# Transport barrier onset and edge turbulence shortfall in fusion plasmas

Guilhem Dif-Pradalier [1 ✉], Philippe Ghendrih[1], Yanick Sarazin [1], Elisabetta Caschera[1], Frédéric Clairet[1], Yann Camenen[2], Peter Donnel [1], Xavier Garbet[1], Virginie Grandgirard[1], Yann Munschy [1], Laure Vermare[3] & Fabien Widmer [2,4,5]

Magnetic confinement fusion offers the promise of sustainable and safe energy production on Earth. Advanced experimental scenarios exploit the fascinating yet uncommon ability of confined plasmas to bifurcate into states of enhanced performance upon application of additional free energy sources. Self-regulation of small-scale turbulent eddies is essential to accessing these improved regimes. However, after several decades, basic principles for these bifurcations are still largely debated and clarifications from first principles lacking. We show here, computed from the primitive kinetic equations, establishment of a state of improved confinement through self-organisation of plasma microturbulence. Our results highlight the critical role of the interface between plasma and material boundaries and demonstrate the importance of propagation of turbulence activity beyond regions of convective drive. These observations strongly suggest a paradigm shift where the magnetised plasma at the onset of enhanced performance self-organises into a globally critical state, 'nonlocally' controlled by fluxes of turbulence activity.

[1] CEA, IRFM, F-13108 Saint-Paul-lez-Durance, France. [2] Aix-Marseille Université, CNRS PIIM, UMR, 7345 Marseille, France. [3] L.P.P. Ecole Polytechnique, Palaiseau, France. [4] International Research Collaboration Center, National Institutes of Natural Sciences, 105-0001 Tokyo, Japan. [5] Max Planck Institute for Plasma Physics, D-85748 Garching, Germany. ✉email: guilhem.dif-pradalier@cea.fr

In the quest for magnetic confinement fusion, field geometry plays an important role. Magnetic field lines in tokamaks or stellarators are built such as to trace out closed toroidal flux surfaces with a high degree of symmetry. Field symmetry is known to bolster confinement, enabling entrapment of the ionised plasma. Symmetry breaking however is common and usually results in net particle, energy, or momentum sinks and ultimately in degradation of confinement. In particular, toroidal symmetry in the plasma core where fusion reactions occur breaks down in the peripheral plasma as flux surfaces open up and field lines intercept material boundaries. The transition from closed to open field lines is usually abrupt, occurs about the so-called magnetic separatrix and plays an important role for confinement.

Established practice oft distinguishes, as sketched in Fig. 1, between a confined 'core' region, dense and hot, an unconfined peripheral boundary layer (the 'Scrape-Off Layer' or 'SOL') and an in-between 'edge' region, loosely defined, set between core and separatrix. The SOL is cold and rarefied; it starts at the magnetic separatrix and is mapped out by the open field lines which connect magnetically to the material boundaries. Core and SOL have been extensively studied, mostly independently, the edge usually serving in modelling as fixed boundary condition for both, its dynamics difficult to apprehend. Strict decoupling however between all three regions is increasingly being questioned. Tokamak plasmas are indeed prone to self-organisation and mounting evidence suggests interplay between core, edge and SOL. Realistic modelling of fusion-grade plasmas must address this delicate balance and tackle dynamics in the vicinity of the closed to open field line interface.

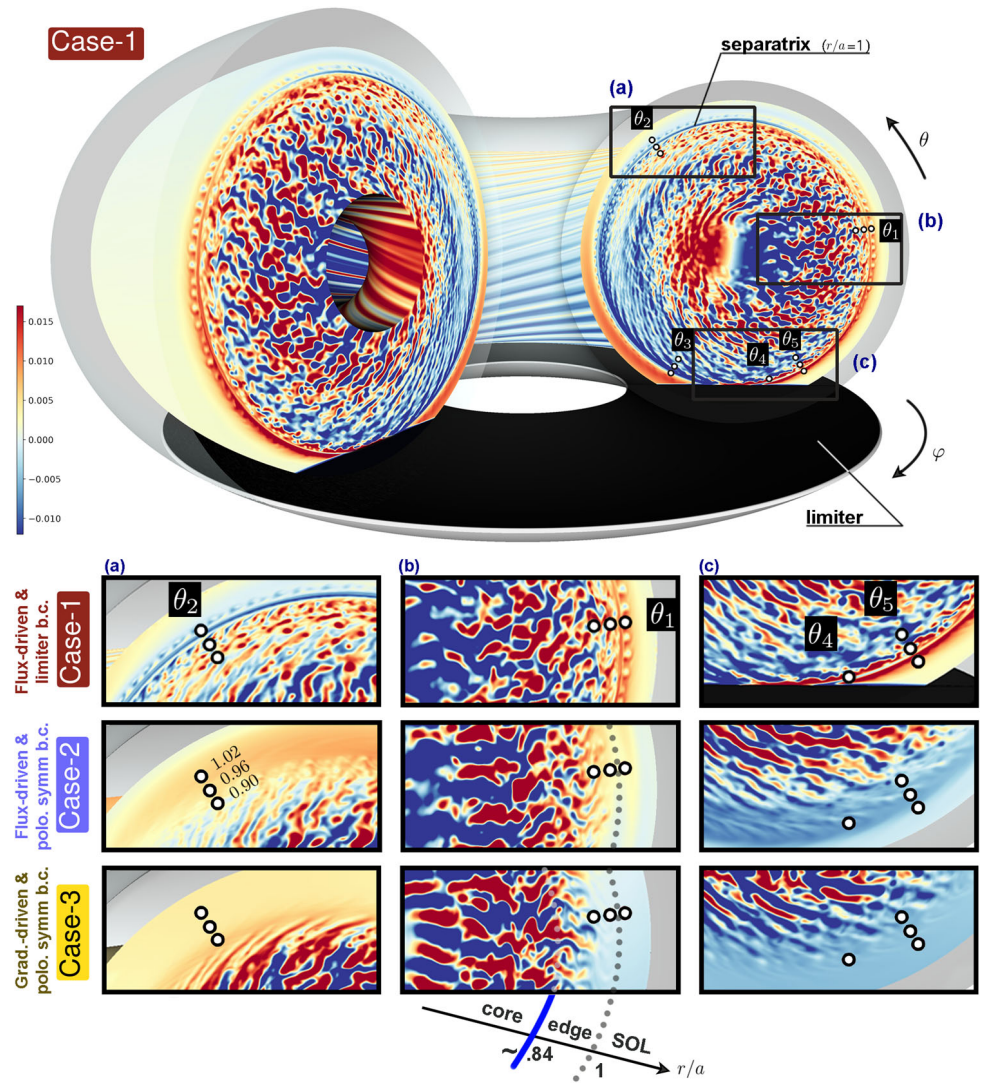

**Fig. 1 Snapshots of the electrostatic potential in three different configurations, at statistical equilibrium.** Case-1 is the reference flux-driven configuration and features, as in experiments, a transition from closed to open field lines in the Scrape-Off Layer (SOL) through introduction of a poloidally localised toroidal limiter (separatrix is at normalised radius $\rho = r/a = 1$) and a wall (at $\rho = 1.3$) within the computational domain. Case-2 differs from Case-1 through its external sink alone: uniformly distributed in the poloidal (along $\theta$) and of gradually increasing strength past $\rho \geq 1$ (see Methods, section "Model equations"). Case-1 and Case-2 are initialised identically; subsequent discrepancies in their temporal evolution is thus direct incidence of the outer boundary condition alone. Case-3 is the gradient-driven twin of Case-2: it tests for the influence of turbulence spreading on the global self-organised state. Background mean gradients for Case-3 are the converged profiles of flux-driven Case-2, other parameters being equal between both computations. Case-3 hovers throughout nonlinear evolution about the statistical state of flux-driven Case-2, effectively prescribing a scale separation between background and fluctuations. For proper comparisons, the same radial-poloidal areas are magnified for each Case: top of the machine [column **a**]; low-field side midplane [column **b**] and bottom regions [column **c**]. Reference (circled) positions are systematically displayed in all panels at various radial ($\rho = 0.9, 0.96$ and $1.02$) and poloidal ($\theta_1$ through $\theta_5$) locations. The separatrix in Case-1 is clearly apparent; for comparison, it is drawn at the same locations for Case-2 and Case-3 as dotted grey lines in **b**. The solid blue line at $r/a \sim 0.84$ corresponds to the transition from (linearly) convectively unstable core to convectively stable edge [see Methods, section "Linear stability analysis" and ref. [37]].

The near-separatrix edge region is a cornerstone of fusion research, where spontaneous transitions from low-confinement "L-mode" to high-confinement "H-mode" occur[1]. The H-mode branch of operation is one of several improved confinement states that have been experimentally discovered, revitalising the fusion programme towards ITER. Significant performance enhancement has been made possible by exploiting the spontaneous L to H transition, induced by increased radial electric field shear Er'. Onset of differential rotation, which scales with Er' and steepening of the ion pressure profile in a localised region of the L-mode edge—the so-called "pedestal"—stabilises turbulence, reduces transport and initiates a self-reinforcing feedback[2,3], which locks-in the bifurcated state. Description of these dynamics from first principles is still lacking. It requires the comprehensive depiction of transport processes in the plasma edge, prior to bifurcations—i.e., in the degraded low-confinement L-mode regime. Our study therefore focusses on this regime and on understanding the early phases of a bifurcation to improved confinement.

Microturbulence dominates the transport processes there, through stress and electric field shear production. We discuss a generic situation, based on experimental parameters where the following conundrum is found: experimentally, the edge is measured to be turbulent, with fluctuations increasing with proximity to the separatrix[4]. In contrast, local analysis of the profiles predicts convective stability in the edge and increased stability with proximity to the separatrix. The edge region would thus appear as unfit to produce or sustain a turbulent state. This opposes the experimental trend and precludes the possibility for turbulence-induced bifurcations to improved confinement. This "shortfall" of predicted turbulence power near the plasma separatrix—see e.g., ref.[5] has been puzzling scientists of the field for decades.

We show, from the primitive kinetic equations, a possible resolution to this problem. Understanding the origin of turbulence activity in the edge requires considering the interplay between closed and open field lines. Magnetic connection to the material boundaries deeply modifies global convective stability at the separatrix. An additional source of free energy arises there, resulting in the confined plasma being driven unstable. We describe whereby fluctuations, initially localised in a narrow peripheral area of the plasma expand beyond their region of convective drive and spread throughout the stable edge. Whilst expanding, turbulent eddies organise such that a transport barrier spontaneously arises and the plasma transitions to a state of improved confinement. We do not claim description of the transition to a fully bifurcated H-mode as this may require additional physical ingredients further discussed below. Rather, there is merit in (relative) simplicity. We highlight a minimal set of robust ingredients, ever-present in the edge of magnetised plasmas, which allows to sustain a convectively unstable edge despite large areas being convectively (linearly) stable. In the process, we discuss the causal chain of events that leads to large flow shear growth at the separatrix. This lays the groundwork for the elucidation of dynamically-pertinent feedback loops in transport barrier onset, highly relevant to the fusion programme.

## Results and discussion

**The plasma edge: a modelling challenge.** A tradition of works has attempted to model the separatrix and edge regions. Owing to the strength of the guiding magnetic field, plasma turbulence is often stated as quasi two-dimensional. Unstable modes helically extend along field lines whilst being at leading order radially pinned to flux surfaces. On the basis of this property, separations of scales between fluctuations and mean background are commonly performed. Such approaches, referred to as "gradient-driven"[6–13] are computationally efficient and have extensively been used in the plasma core. Their validity however faces challenges in the edge and wanes further outwards whilst nearing the separatrix and SOL. Turbulent scales at the plasma edge indeed become comparable to free energy gradient scales and oft-assumed separations between a slow, large-scale background and fast, small-scale fluctuations become inadequate, as already noted by Kadomtsev[14].

In the peripheral plasma, gradients are steep, with intrinsic temporal variability. Precise experimental measurements are challenging. Fluxes may vary by factors in gradient-driven frameworks upon scanning imposed mean gradients within experimental error bars. This generates considerable difficulty to predict performance of magnetic confinement devices or to assess safe operation of ITER. At statistical equilibrium, however, power that has been imparted to the plasma must come out. This driving power is known. This has led to proposing a paradigm shift in modelling where known, imposed fluxes drive the system and both mean gradients and fluctuations dynamically evolve in concert, act and back-react on one another. Such a framework is referred to as "flux-driven"[15–22]: its added feature is to probe the multiscale interplay between instability micro-scales and meso- to macro-scales. As it relaxes assumptions of scale separation and comes at the price of a significant (tenfold or more) increase in computational demands.

In fully bifurcated H modes, the large gradients have built up over a limited region of space—the so-called edge "pedestal". With such large gradients $1/L_p$, electromagnetic modes may become more prominent as the effective beta $\beta_{eff} = \beta(qR/L_p)^2$ for instability increases. Validity of gyrokinetics[23], which requires $\rho/L_p < 1$ is also sometimes questioned. No understanding of H-mode however may be complete without comprehensive understanding of the degraded confinement in the first place. The present study is thus concerned with elucidating onset of edge dynamics in L-mode where $\beta_{eff}$ remains moderate and where millimetric Larmor radii and centimetric gradient lengths is firm ground for gyrokinetics. We further note that several attempts have tried to explain the "edge shortfall" as a consequence of missing electromagnetic modes or breakdown of the gyrokinetic assumption[24]. To our best knowledge, all such attempts, performed using local gradient-driven approximations, have repeatedly found a transport shortfall at nominal edge parameters.

Given the complexity of understanding early stages of edge pedestal build-up, there is therefore considerable merit to search for a "bare-bones first principles" approach as the model appropriate to provide resolution of the edge transport shortfall problem. The present work is based on flux-driven electrostatic gyrokinetics, including transition from closed to open field lines and proposes a clear step in this direction. It takes a complementary approach to the common flux-tube, local or even global gradient-driven approximations. Massive increase in computational power in recent years has been instrumental to permit the present investigation. Our study connects to a larger tradition of works that have shown pervasive though usually highly non-trivial impact of boundary conditions and of forcing on nonlinear evolution of complex systems. Recent examples in fluids have experimentally demonstrated that conjugate forcings (either energy-constrained or temperature-constrained) produce non-unique statistical steady states as well as different bifurcations between regimes (e.g., ref.[25]). We highlight similar conclusions here for magnetic fusion, with substantial implications for modelling.

To illustrate these points, we systematically compare a series of three high-resolution numerical experiments (Cases 1 through 3

in Fig. 1). The coupled gyrokinetic and Poisson equations are time stepped in five dimensions, from core ($\rho = r/a = 0$) to wall ($\rho = 1.3$), using the gyrokinetic GYSELA framework[26]. Results presented here are reproducible. Besides, GYSELA has been extensively benchmarked over the years against standard tests and verified against other comparable codes. Plasma parameters of the Tore Supra shot #45511 are mimicked and in particular plasma-boundary interaction is modelled through introduction within the computational domain of a wall and a toroidal limiter, as in the experiment (see Methods, sections "Model equations" and "Physical parameters and robustness"). The limiter is a key plasma facing component, toroidally symmetric (alongside $\varphi$, Fig. 1) and poloidally localised (about $\theta = 3\pi/2$) which has been extensively used to handle power loads in fusion devices. At its central location, plasma and material boundary connect tangentially at normalised radius $r/a = 1$. This defines the magnetic separatrix and the transition between closed, confining magnetic surfaces and open field lines.

'Case-1' is our reference computation. It employs a flux-driven framework in a realistic limiter and wall geometry. Case-2 and Case-3 are companion computations that only differ from Case-1 either by forcing (flux-driven versus gradient-driven) or by boundary conditions (limited versus poloidally symmetric). The goal in investigating these two boundary conditions is to elucidate the role of poloidal asymmetry and of closed/open field line transition in edge barrier onset. 'Case-2' is thus the flux-driven twin of Case-1 but without limiter and wall to assess impact of the closed to open field lines transition. Case-2 features commonly used poloidally symmetric (uniform along $\theta$) and radially progressive absorbent boundary conditions. The goal in investigating forcing is to provide a comprehensive examination of the actual role of turbulence spreading[27–29] in edge self-organisation.

Spreading is generic in fluids and plasmas and refers to the possibility for turbulence to expand beyond its region of convective drive. This property is sometimes also referred to as nibbling[30] or engulfment[31] in jet interfaces, overshooting, or penetration[32] in geophysical and astrophysical fluids. It has played an important role to explain subcritical transitions to turbulence in parallel flows, which pertain to the class of directed percolation[33]. Spreading is strongly appealing to explain how the L-mode edge may sustain a turbulent activity[34] despite often being predicted to be either convectively stable (next section) or marginally unstable.

Quantification of the actual importance of spreading from primitive equations is largely uncharted. Precise comparison between flux-driven and gradient-driven forcing addresses this question as the scale separation in gradient-driven frameworks hinders the back-reaction of fluctuations on the driving mean gradients, an essential ingredient for spreading. Case-3 thus shares with Case-2 the same symmetric boundary conditions whilst being its gradient-driven analogue. Comparison of Case-2 to Case-1 and Case-3, therefore, provides a comprehensive view of the role of spreading, comprehensively discussed in section "Turbulence spreading: instrumental to edge turbulence build-up".

**Free energy injection at the magnetic separatrix.** Interestingly, a marked difference in edge turbulence (Fig. 1) is found upon modification of the outer plasma-boundary conditions from limiter (poloidally asymmetric with closed/open field line transition) to radially progressive and poloidally symmetric (closed field lines only). Gradient inhomogeneities in the poloidal plane are pervasive in experiments, though often elusive to diagnose and may stem from the presence of a limiter, of a radiating X point or of high-Z impurities. Their determination is a central matter for transport[35,36]. Interaction with the limiter in Case-1 triggers a spontaneous symmetry breaking: it onsets and sustains a density shelf in the poloidal vicinity of the limiter [Fig. 2a, c], a poloidal distribution of temperature gradients [Fig. 2b, d] and also induces a radial electric field well (Fig. 2f), a signature of improved confinement. All three features are absent in poloidally homogeneous Case-2 and Case-3. Profiles in panels (a) through (d) are colour-coded with poloidal proximity to the limiter (e).

To clarify the nature of the edge free energy in all three Cases, extensive linear stability analysis has been performed using the initial value framework of the Gyrokinetic Workshop (GKW) code[12] at various radial-poloidal ($\rho_j$, $\theta_k$) edge locations (circles in Fig. 1). It leads to the following picture [Methods, section "Linear stability analysis" and ref. [37]]: (i) edge profiles in Case-2 and Case-3 are linearly stable to drift-wave, interchange, and parallel shear flow instabilities past $\rho \gtrsim 0.9$ whilst (ii) a localised region of instability, dominantly of interchange character appears in Case-1 immediately inside the separatrix due to the presence of the limiter, on both sides of it. This large portion $\rho \gtrsim 0.9$ of the edge which is predicted as convectively stable is sometimes called the "No Man's Land" (NMsL), bearing witness to its puzzling stability.

Importantly, mean profiles at the low-field side midplane ($\theta = 0$) where experimental measurements are usually performed are very similar between all three Cases. On the basis of local analysis of available free energy, undistinguishable nonlinear evolution would be predicted there between all three Cases. The limiter modifies this picture. As a cold spot, it tends to create radial-poloidal pressure anisotropies in its vicinity, in a manner akin to thermodiffusion described by off-diagonal (density and momentum pinch) terms in the transport matrix. Such a mechanism provides robust free energy injection nearby a localised heat or momentum sink from combined onset of a radial and poloidal pressure gradient.

The main findings are as follow: plasma-boundary interaction provides a pathway to a novel source of free energy that locally builds ion-scale, electrostatic turbulence in the vicinity of the cold sink. Then turbulence self-advection ("spreading"), which results from the possibility of flux-driven micro- to meso-scale interplay, redistributes fluctuations globally despite underlying convective stability. We note that this convective stability is robust and independent of the details of the electron response, Boltzmann or not (the edge $r/a > 0.84$ is linearly stable, with or without this approximation, as shown by GKW). This is a possible indication that the nature of the instability is likely not as critical to edge dynamics as is the combination of plasma-solid interactions and turbulence spreading. Lastly, we find that curing the edge shortfall problem through combination of the above mechanisms naturally opens the route for pedestal build-up, thus highlighting intimate connection between the edge shortfall problem and access to enhanced regimes of confinement. Vorticity advection appears as the important causal link to reconcile the somewhat counter-intuitive idea that additional turbulence in the edge may lead to pedestal build-up.

**Turbulence spreading: instrumental to edge turbulence build-up.** Self-advection (spreading) of patches of turbulence intensity $\mathfrak{I}$, may be quantitatively followed through wave-energy budget[38]. The relevant conserved quantity is the negative of the entropy density ($n\mathfrak{I}$), which involves the ambient density profile $n$ and naturally leads to amplification of fluctuations $\mathfrak{I}$, as $n$ decreases in the plasma edge. Wave-energy density conservation indeed implies that the fewer the oscillators, the larger the oscillations. The related conservation equation[39]: $\frac{D}{Dt}(n\mathfrak{I}) + \nabla \cdot \Gamma_S = \mathcal{S}$ features the energy flux $\Gamma_S(r, \theta, t) = \frac{1}{2} \langle \int d^3v (\mathbf{v}_E \cdot \nabla r) \frac{(\delta f)^2}{F_M} \rangle_\varphi$ as kinetic proxy for spatial turbulence spreading. Here $D/Dt$ denotes the turbulent convective derivative, $\mathcal{S}$ free energy injection and dissipation mechanisms, $(\delta f)$ the departure of the full ion distribution function $\bar{F}_s$ [whose evolution satisfies Eqs.(2)–(6), see Methods] from an ensemble averaged Maxwellian $F_M$ reconstructed from the evolving local density and temperature profiles, $(\mathbf{v}_E \cdot \nabla r)$ is the radial $\mathbf{E} \times \mathbf{B}$ velocity and $\langle \cdot \rangle_\varphi$ toroidal averaging.

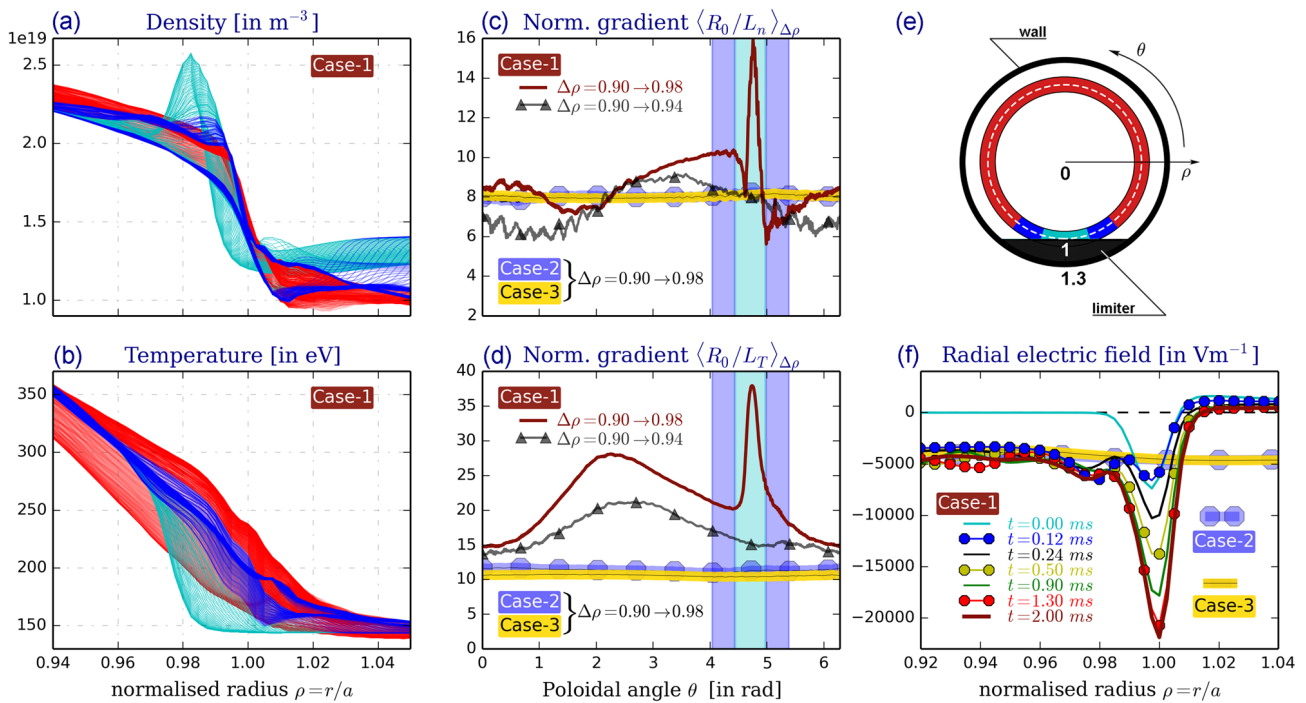

**Fig. 2 A novel source of free energy arises from spontaneous symmetry breaking at the plasma-boundary interface.** The radial profiles (along $\rho$) of density (**a**), temperature (**b**) and the poloidal profiles (along $\theta$) of their respective normalised gradients (**c, d**) are colour-coded with poloidal proximity to the limiter (**e**). The cyan [resp. dark blue and red] profiles in **a**, **b** represent the radial profiles at different poloidal angles in the cyan [resp. dark blue and red] area of **e**. The poloidal profiles of radially averaged normalised gradients are shown in **c, d**. The solid red [resp. grey triangles] line is radially averaged over interval $0.90 \leq r/a \leq 0.98$ (resp. over $0.90 \leq r/a \leq 0.94$). Similarly, the cyan [resp. dark blue] vertical shades correspond to the cyan [resp. dark blue] area of **e**. Hot plasma particles rapidly stream towards the limiter along field lines as well as through the action of vertical $\mathbf{B} \times \nabla B$ magnetic drift (here pointing towards the limiter). This dynamics leads in Case-1 to the onset and sustainment of an over-dense poloidal density shelf (**a, c**) near the limiter as well as a localised temperature gradient (**b, d**) and a radial electric field well, poloidally averaged (**f**), absent in poloidally homogeneous Cases 2 and 3 (thick dotted light purple and thick solid yellow respectively). This poloidal anisotropy is especially significant within 5% of the separatrix, as seen through radial averaging $\langle \cdot \rangle_{\Delta \rho}$ of equilibrium mean gradients (**c, d**). Gradient anisotropy decreases in magnitude in the near SOL and poloidally farther from the limiter. Large and anisotropic equilibrium mean gradients in the limited configuration result in locally-enhanced free energy sources for the turbulence in the outermost 5% layer of the confined edge. Inversely, Case-2 and Case-3 display uniform and moderate mean gradients, shown in thick yellow and dotted purple.

The "beach effect" model[38] does not satisfactorily account for our observations: core waves propagating on large radial distances and amplifying via convective instability in the edge is not the main mechanism found here. The maximum of relative fluctuations $\delta n/n$ [Figs. 3i and 4c] is also around the top of the high gradient region. This model however provides an interesting mental framework as it emphasises the role of poloidal asymmetry of propagating waves in order to obtain in toroidal geometry a non-zero radial energy flux, i.e., spreading, which we show below to be an essential ingredient.

Self-advection of turbulent patches is best apprehended as $\Gamma_S$ increments. Fig. 3 displays times series of poloidal cross-sections of spreading increments $\Delta S = \Gamma_S(r, \theta, t) - \Gamma_S(r, \theta, t_{ref})$ for Case-1. Three phases appear: a NMsL devoid of fluctuations (Phase I, subplots Fig. 3a through 3d) is clearly visible in the early stages of Case-1 for all poloidal angles and for $0.85 \leq r/a \leq 1$, echoing aforementioned linear stability of underlying profiles. Strong and persistent inward advection of turbulence intensity originates at the edge-limiter boundary ($\sim \theta = -\pi/2$), propagates (white arrows) radially inwards and poloidally anticlockwise all the way to the top ($\theta = +\pi/2$), in $\sim 0.1$ ms [$16,700\,\Omega_{ci}^{-1}$]. This turbulence intensity front propagates inwards until about $r/a = 0.82$, which amounts to a radial penetration depth of about 60 local Larmor radii, i.e., 10–12 local turbulence correlation lengths.

From there [Phase II, Fig. 3e though 3f], outwards radial spreading (black arrows) accompanied by clockwise poloidal motion from the plasma high field side and top regions to the low-field side midplane ($\theta = 0$) fills-in NMsL with turbulence in about 0.5ms. This to and fro redistribution of turbulence intensity bridges the region of free energy injection near the limiter with the upstream confined core. As both regions connect, core turbulence spills over, further enhancing edge fluctuation levels. This complex radial-poloidal dynamics finds an echo in the synthetic reconstructions [Fig. 3h, i] of the radial profiles of turbulent fluctuations $\delta n/n$ around $\theta = 0$, as would be measured in Tore Supra. These synthetic profiles are plotted against typical actual measurements using fast-swept reflectometry (shaded grey)[40].

A dynamic equilibrium (Phase III, Fig. 3g), characterised by quasi-periodic relaxations of the edge turbulence is reached at later times (from 1.55 ms onward). Bursts of clockwise rotating outgoing patches of turbulence (black arrows in Fig. 3g, i) equilibrate incoming anticlockwise limiter-borne fluxes of turbulence intensity.

Comparatively, Case-2 [Fig. 4a] and Case-3 [Fig. 4b] equilibrate on faster timescales, with spreading patterns dominantly outwards and reaching equilibrium in ~0.6 ms. A clear shortfall is observed [Fig. 4c, top] past $0.85 \leq r/a$ for Case-2 and past $0.8 \leq r/a$ for Case-3, again reminiscent of linear stability of underlying gradients, uncompensated by fluxes of turbulence activity from distant active (core or boundary) regions. Spreading in gradient-driven Case-3 is modest, at most about a few local turbulence correlation lengths, as in earlier works[28], and insufficient to explain the required fluctuation levels both in the deep core and in the edge.

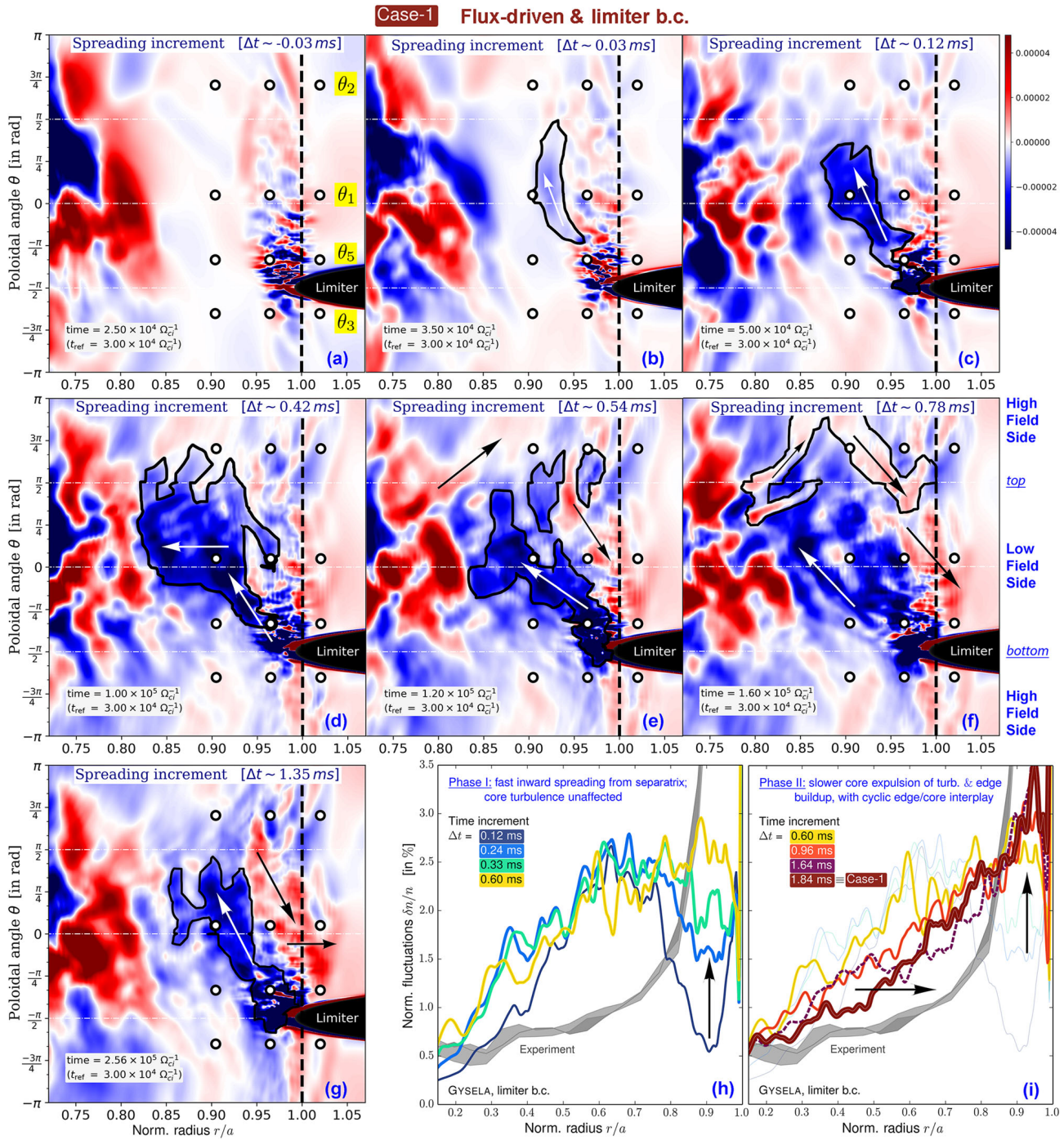

**Fig. 3 Spatiotemporal redistribution of induced free energy injection at the plasma-boundary interface (for flux-driven Case-1). a–g** times series of poloidal cross-sections of spreading increments $\Delta S$. Positive (red) increments represent radially-outward fluxes of turbulence intensity; inward fluxes of turbulence intensity are pictured as negative (blue) increments. The limiter is responsible for vigorous mixing in its immediate poloidal (between $\theta_3$ and $\theta_5$) and radial ($0.95 \leq r/a \leq 1.0$) vicinity. The drive endures throughout linear (**a**) and nonlinear evolution [**b** through **g**]. The choice of an early reference time $t_{ref} = 30,000 \, \Omega_{ci}^{-1}$ allows to follow the full unfolding of the spreading sequence in the edge. With a later choice for $t_{ref}$ the early spreading sequence would not appear as clearly; later nonlinear dynamics (Phases II and III [$\Delta t \geq 0.6$ ms]) on the other hand would be qualitatively unchanged: cyclic equilibration on the outboard midplane between incoming and outgoing fluxes of turbulence intensity that originate from distinct poloidal regions. Inward spreading (Phase I, **h**) is responsible for edge increase of turbulence activity at the outboard midplane. As 'No Man's Land' (the region $0.85 \leq r/a \leq 1.0$) fills-in, outgoing turbulence activity spreads to the edge [Phase II, **i**], further enhancing relative fluctuation levels in the outer edge. At equilibrium (Phase III, **g**), the fluctuation profile hovers about its 1.84 ms value (**i**) equilibrating incoming and outgoing spreading increments.

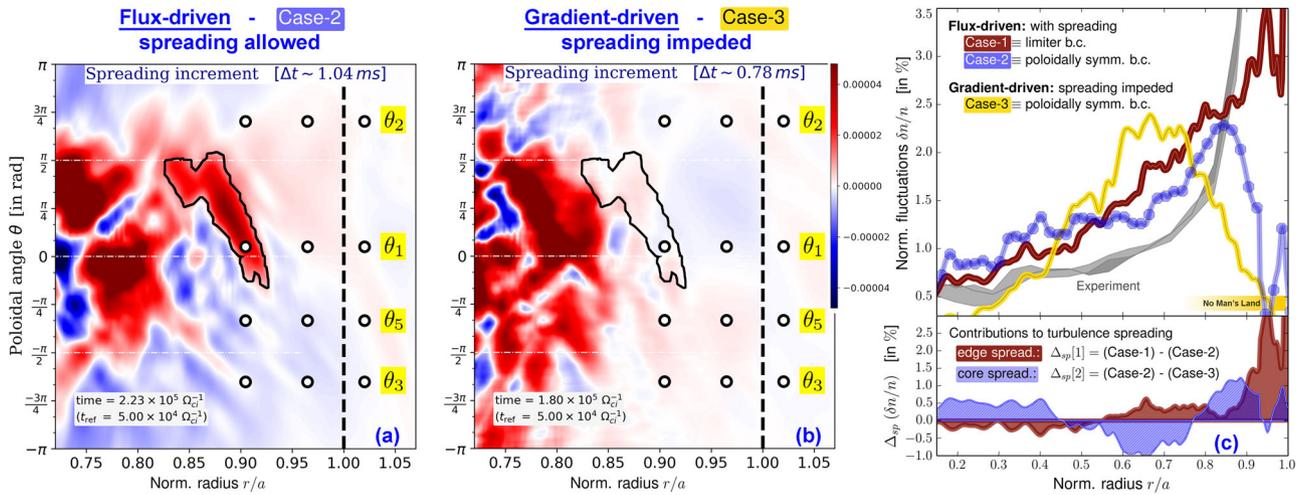

**Fig. 4 How forcing and interplay with boundaries impact redistribution of turbulence activity and globally affect transport.** Spreading increments for Case-2 (**a**) and Case-3 (**b**) at time $t_{ref} = 50,000\,\Omega_{ci}^{-1}$ when both Cases enter nonlinear regime in the edge. A clear shortfall of turbulence activity is visible at statistical equilibrium (**c** top) for both Case-2 (purple) and Case-3 (yellow) with respect to experimental (grey) and Case-1 (red) fluctuation profiles $\delta n/n$. 'No Man's Land' (NMsL) is the region where such shortfalls occur. NMsL, when comparing Case-2 to Case-1 approximately extends between $0.85 \leq r/a \leq 1.0$. When comparing Case-3 to Case-1, NMsL approximately extends between $0.8 \leq r/a \leq 1.0$. Differences $\Delta_{sp}[1]$ and $\Delta_{sp}[2]$, respectively quantify (**c** bottom) the importance of turbulent spreading of fluctuations from separatrix ➡ core and from core ➡ edge.

Normalised contributions $\Delta_{sp}[j] = \delta n/n[j] - \delta n/n[j+1]$ further quantify the weight of spreading [Fig. 4c, bottom] at statistical equilibrium, $j$ denoting the Case index. Quantity $\Delta_{sp}[1]$ assesses outside ➡ in spreading. Clearly, near-separatrix ➡ core contamination of limiter-borne turbulence activity accounts almost in full for the fluctuation dynamics in the outer radial 10% of the confined plasma. Similarly, since Cases 2 and 3 have the same outward boundary conditions, $\Delta_{sp}[2]$ quantifies the weight of both core fluctuation redistribution and inside ➡ out spreading of core turbulence intensity towards the edge, amplified by the beach effect.

Mean gradients in Case-2 and Case-3 being the same by construction, the reversing sign of $\Delta_{sp}[2]$ illustrates the natural tendency of the flux-driven system with respect to the gradient-driven framework to radially redistribute patches of turbulence intensity, both outward ($0.8 \leq r/a \leq 0.92$) and inward ($r/a \leq 0.4$). Extra spreading in flux-driven approaches, illustrated in Fig. 4a by the black contour about $\theta_1$ and lack thereof in Case-3 [same contour, Fig. 4b] explains the gradient-driven underprediction of fluctuation levels in the deep core $r/a \leq 0.45$ and outer edge regions $0.85 \leq r/a \leq 0.90$ as well as its overprediction in the intermediate linearly-unstable region $0.55 \leq r/a \leq 0.75$.[41]

**Causality in transport barrier onset.** We have established that the combination of free energy injection at the plasma-boundary interface and turbulence spreading provides a robust pathway to a turbulent edge. The last part of the triptych is to substantiate how turbulent eddies organise so as to unlock access to improved confinement. A fecund way to address this problem is to examine the causal chain of events that presides over edge transport barrier onset.

Causality is actively debated in information theory[42,43]. We focus here on a nonlinear extension of the Granger causality, using the information-theoretic "Transfer Entropy" (TE) approach[44,45] to question the causal chain of events underpinning transport barrier build-up. TE is generic and allows for a diagrammatic representation of net (directional) information transfer in bivariate time series analysis. We apply it[41] to the spontaneous emergence of the stable and localised transport barrier which develops at the closed/open field line transition through limiter-induced symmetry breaking. The sharp radial electric field well at the plasma edge [Fig. 2f] is clear signature of it. Application of TE on electric field data from the primitive flux-driven equations opens captivating possibilities to shed light on the basic mechanisms at play.

Radial electric field dynamics satisfies a vorticity balance, Eqs. (7–11) and refs. [46–48] which may be systematically derived from the primitive gyrokinetic-quasineutrality equations, Eqs. (2–6). All terms in Eqs. (7) and (8) are computed and those with significant magnitudes are displayed in Fig. 5a. Overall vorticity balance is precisely satisfied (0.9%), which allows to write at leading order:

$$\partial_t \langle \Omega_r \rangle + \frac{1}{r}\partial_r r \left[ \langle v_{Er}\Omega_r \rangle + \langle v_{\star r}\Omega_r \rangle - \left\langle v_{\star \theta} \frac{1}{r}\partial_\theta E_r \right\rangle \right] = r.h.s \approx 0 \quad (1)$$

where $\langle \cdot \rangle$ denotes toroidal averaging.

Radial vorticity $\langle \Omega_r \rangle$ (flow shear) thus dominantly evolves through the combined influence of three fluxes: the usual $\mathbf{E} \times \mathbf{B}$ (radial) advection of vorticity $\langle v_{Er}\Omega_r \rangle$ [hereafter denoted "Reynolds force"] and two seldom discussed mechanisms: diamagnetic (radial) advection of vorticity $\langle v_{\star r}\Omega_r \rangle$ and diamagnetic (poloidal) advection of poloidal inhomogeneities of the radial electric field $\langle v_{\star \theta}\frac{1}{r}\partial_\theta E_r \rangle$ [hereafter, "field advection"]. Unexpectedly, in the early stages of radial electric field build-up both latter contributions display magnitudes larger than that of the Reynolds force [Fig. 5a]. Dynamical significance to vorticity build-up however does not straightforwardly follow. Systematic TE computations for all pairs of terms in Eq. (1) is performed and the relevant interactions are portrayed in Fig. 5c–e.

Area ❺ is the production region for edge fluctuations and naturally sustains local inhomogeneities in the radial-poloidal plane. Larmor motion in this inhomogeneous background is responsible for persistent diamagnetic currents, whose importance is not commonly stressed. Interestingly here: (i) diamagnetic (radial) advection of vorticity $\langle v_{\star r}\Omega_r \rangle$ proves to be at barrier inception the dominant causal agent [Fig. 5d] which directly generates vorticity $\langle \Omega_r \rangle$ and forecasts the Reynolds force $\langle v_{Er}\Omega_r \rangle$.

This central role of $\langle v_{\star r}\Omega_r \rangle$ (ii) endures at later times [Fig. 5e] even when the Reynolds force (towards which $\langle v_{\star r}\Omega_r \rangle$ keeps transferring information) starts contributing more to vorticity production. The oft-expected dominant transfer mechanism: (iii) Reynolds force causing vorticity production becomes major only at later times $\Delta t_2$ and in area ❻ [Fig. 5c], naturally stable without limiter. Area ❻ is stable at earlier times $\Delta t_1$ [Fig. 5b] and remains such until turbulent fluctuations have spread there ('downstream') from production region ❺.

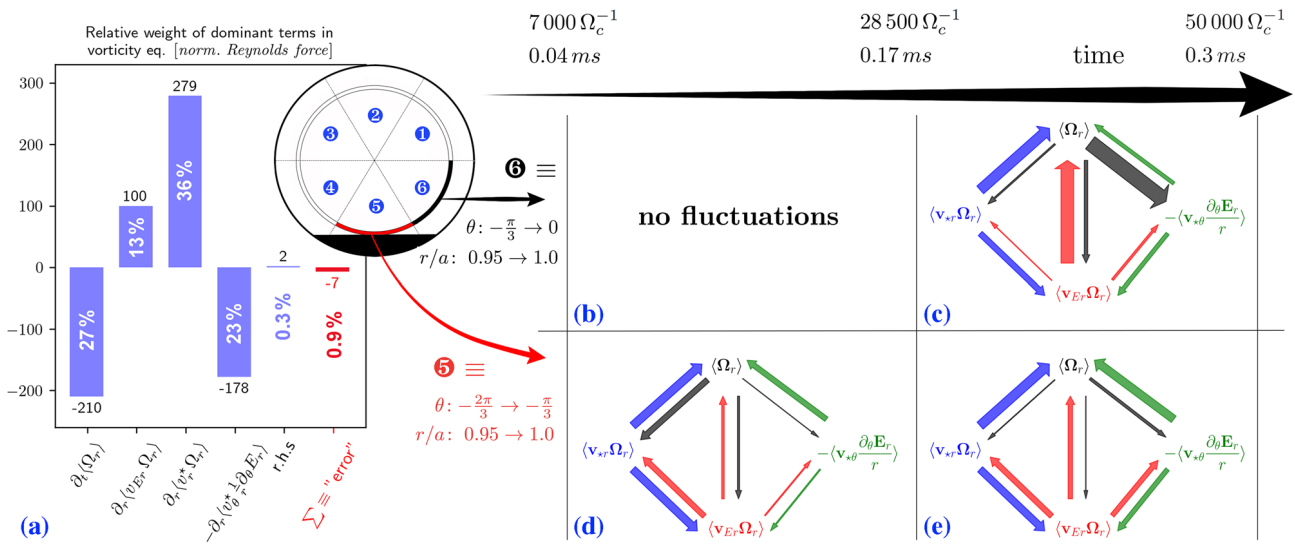

**Fig. 5 Accurate vorticity balance allows investigation of causality in transport barrier build-up.** Causal interactions are investigated in the early stages of barrier build-up (initial 0.3ms, or 50, 000 $\Omega_{ci}^{-1}$), between $0.95 \leq r/a \leq 1.0$ where limiter-induced fluctuations are borne. (**a**): Precise test of vorticity balance through systematic computation of all terms in Eqs. (7) and (8). Their relative magnitudes are estimated through radial ($0.95 \leq r/a \leq 1.0$), poloidal ($0 \leq \theta \leq 2\pi$), toroidal ($0 \leq \varphi \leq \pi/5$) and temporal ($0 \leq t\Omega_{ci} \leq 50,000$) averaging. Their algebraic sum is displayed as "error" (value $-7$ corresponds to 0.9% error). All terms in Eq. (8) are individually small, with r.h.s denoting their sum. (**b** through **e**): the outer corona ($0.95 \leq r/a \leq 1.0$) of plasma volume is divided in six areas; we especially focus on areas ❺ ($-2\pi/3 \leq \theta \leq -\pi/3$), centred about the limiter where the radial electric field well originates and area ❻ ($-\pi/3 \leq \theta \leq 0$) into which turbulence later spreads. Dominant information transfer from systematic pairwise TE computations is diagrammatically represented for areas ❺ and ❻, during two time intervals. During $\Delta t_1$ (0.04 ms to 0.17 ms), turbulence is confined to area ❺ (**d**) and other regions are stable (**b**). During $\Delta t_2$ (0.17 ms to 0.3 ms), fluctuations have contaminated area ❻ (**c**) whilst keep being produced in area ❺ (**e**). Arrows indicate the direction of information flow and their respective thickness is proportional to the amount of information actually transferred.

Our results may seem in apparent contrast with earlier studies, e.g., ref. [49] which has emphasised the Reynolds force as the main flow driving mechanism. This may possibly be only in appearance. Indeed, measurements in ref. [49] are performed in slab, poloidally symmetric configurations. This effectively minimises the $\langle v_{\star r} \Omega_r \rangle$ contribution with respect to our results which clearly highlight the role of poloidal asymmetry and pressure inhomogeneities in toroidal geometry. Furthermore, in the present manuscript we resolve the short time dynamics of radial electric field growth from a zero state, whilst experimental measurements are in steady-state and temporally coarser grained. Interestingly, at later times, in established steady-state and far from the limiter, our results (not displayed here) tend to show that contribution of the "usual" Reynolds force to vorticity balance becomes more significant. This may interestingly illustrate a discrepancy between mechanisms that trigger barrier onset (for which radial diamagnetic advection of vorticity is central) from mechanisms which sustain it (there, Reynolds force is an important player together with diamagnetic advection of vorticity which remains dynamically significant).

Let us also note that $\langle v_{\star r} \Omega_r \rangle$ and $\langle v_{Er} \Omega_r \rangle$ are often expected to counteract each other. This is however not generally the case in flux-driven regimes, as shown in ref. [48]. Compressibility matters and the transverse pressure is not found to act as a passive scalar —it is not merely advected by the $\mathbf{E} \times \mathbf{B}$ flow. As a result, finite Larmor radius (FLR) corrections to the Reynolds stress tensor cannot be ignored in a flux-driven setting, and may even prove dominant in magnitude.

TE analysis also highlights another (new to our knowledge) mechanism for both vorticity production and shear dissemination: (poloidal) diamagnetic currents contribute to vorticity build-up whilst poloidally propagating radial electric field poloidal inhomogeneities $\langle v_{\star\theta} \frac{1}{r} \partial_\theta E_r \rangle$ [Fig. 5d, e]. Farther downstream [area ❻, Fig. 5c] vortical structures $\langle \Omega_r \rangle$ transfer information to field advection; this mechanism contributes to expanding the new-sprung $E_r$ well, initially a localised feature of area ❺ and making it a poloidally global feature spanning regions ❶ through ❻.

Causal analysis of transport barrier build-up emphasises the central and somewhat unexpected role of diamagnetic flows ($v_{\star\theta} = \partial_r p_\perp$ and $v_{\star r} = -\partial_\theta p_\perp/r$), shedding light on new or low-keyed mechanisms. This underlines the role of pressure inhomogeneities and FLR effects in barrier build-up. The importance of the latter could have interesting side effects, foremost on I-mode or H-mode accessibility as different isotopes or different classes of particles (electrons, main ions, energetic particles, or impurities) may thus differently contribute to vorticity (shear) production. Provided diamagnetic flows (pressure inhomogeneities) are as important experimentally as they appear in our current study, this could contribute to explaining how different plasma contents may display different thresholds to access bifurcated states of enhanced confinement.

### Conclusions

In the early days of fusion research, where plasma and material boundaries would meet was not the focus of specific attention. This situation has evolved. In part because of the requirement to maintain fluxes of heat or mass incident on the material boundaries at manageable levels and with controlled deposition patterns. In part as well because of the experimental discovery of transport bifurcations, chiefly dependant on the organisation of the peripheral turbulent plasma. This latter question is the focus of our present work.

What we establish is threefold: (i) plasma-boundary interaction deeply modifies convective stability next to the magnetic separatrix. (ii) Resulting locally-borne eddies spread out and destabilise distant regions of the edge and core. A globally organised state emerges, 'nonlocally'[50,51] controlled by fluxes of turbulence activity. (iii) Flow shear builds as eddies (vorticity) are advected, primarily through pressure inhomogeneities. The expanding interface organises into a stable peripheral transport barrier, i.e. into improved confinement, spontaneously.

Mechanism (i) above is generic and likely applies beyond limiter configurations to physical cases with radiation near a magnetic X point (a magnetic saddle point). Direct interplay between confined plasma and material boundaries is reduced in X point configurations. Neutral particles thus certainly play an essential role there as they provide localised dissipation in the vicinity of the X point. An additional source of free energy, linked to the penetration depth of neutrals and akin to the free energy source described here for a limiter can thus be expected in X point configurations. Interestingly, where sharp gradients and a radial electric field well are expected from experiments, state-of-the-art gyrokinetic modelling of ASDEX Upgrade plasmas in X point geometry tend to show without plasma-neutrals interplay smoothly varying profiles of density, temperature and radial electric field across the separatrix, at nominal parameters[52]. This discrepancy may further hint at the central importance of poloidally localised free energy injection near the separatrix, as discussed here, to initiate edge turbulent dynamics, later made global through turbulence spreading (ii). The present argument however does not alone provide an explanation for edge turbulence growth and subsequent confinement improvement in the case where the X point is not dissipative and the radiation front located further out in the SOL, near the strike points. This case should be carefully considered.

Self-advection (spreading) of turbulent fluctuations (ii) is shown to play a central role in the global equilibration of edge turbulence. The connection between spreading and confinement is complex. On the one hand, spreading contributes to disseminating turbulence activity, spatially, so it is usually thought of as detrimental to confinement. On the other hand, the same dynamics that propagates pressure inhomogeneities is shown to contribute (iii) to vorticity fluxes, which result in shear production and are beneficial to confinement. Turbulence spreading and barrier formation are not mutually exclusive; the former is here an important player in the onset of the latter.

This possibility has important implications for modelling. It certainly nudges towards flux-driven frameworks. It also provides a different view on transport barrier formation from a bath of active turbulence. Much focus has indeed been given to understanding the *origin* of the fluctuations (from which instability do they stem: electrostatic, electromagnetic, at ion or electron scales, at which exact location, etc.). The rationale being that the nature of the instability carries over nonlinearly and may critically influence the route towards improved confinement.

Our results suggest that a complementary perspective could prove fecund. Provided that a source of instability exists, especially as shown here near a cold dissipative region, eddies will be spawned and turbulent activity will spread. These properties are robust and a priori independent of the nature of the underlying instability. The relevant then questions become: given a distribution of sources (external heating, recycling,…) and sinks (collisional dissipation, interaction with the boundaries) under which conditions is free energy production sustained? How do localised sources of free energy spread in phase space? What parameters control the branching ratio between flow shear reinforcement and dissemination of turbulence activity? First steps have been provided here and should be expanded as a comprehensive understanding of these questions would likely reshape our understanding of confinement bifurcations, yet a central question for fusion research.

## Methods

**Model equations**. Low-frequency microturbulence in weakly collisional magnetised plasmas is appropriately described within the gyrokinetic framework[23]. The GYSELA code[26] solves the governing coupled gyrokinetic:

$$B_{\|s}^* \frac{\partial \bar{F}_s}{\partial t} + \nabla \cdot \left( B_{\|s}^* \frac{d\mathbf{x}_G}{dt} \bar{F}_s \right) + \frac{\partial}{\partial v_{G\|}} \left( B_{\|s}^* \frac{dv_{G\|}}{dt} \bar{F}_s \right) = B_{\|s}^*(\text{rhs}) \quad (2)$$

$$\text{rhs} = \mathcal{C} + \mathcal{S} + \mathcal{D} - \nu \mathcal{M}^{\text{mat}}(r,\theta)[\bar{F}_s - \mathbb{G}_{\text{cold}}] - \gamma^{\kappa}\left[\bar{F}_s - \mathbb{F}_{F-D}\left(1 + \frac{\langle \bar{F}_s - \mathbb{F}_{F-D}\rangle}{\langle \mathbb{F}_{F-D}\rangle}\right)\right] \quad (3)$$

and quasineutrality:

$$\Delta n_e = \rho + \frac{1}{n_{e_0}} \sum_s Z_s \nabla_\perp \cdot \left(\frac{n_{s_0}}{B\Omega_s}\nabla_\perp \phi\right) \quad (4)$$

$$\frac{T_e}{e}\Delta n_e = \phi - \lambda\left[1 - \mathcal{M}^{\text{SOL}}(r)\right]\langle\phi\rangle_{\text{FS}} - \left[\mathcal{M}^{\text{mat}}(r,\theta) - \mathcal{M}^{\text{wall}}(r)\right]\phi^{\text{bias}} \\ - \lambda\Lambda\left[\mathcal{M}^{\text{SOL}}(r) - \mathcal{M}^{\text{mat}}(r,\theta)\right](T_e - T_e^{\text{b.c.}}) \quad (5)$$

equations for the guiding-centre distribution function $\bar{F}_s$ of ion species $s$, evolved with no separation between equilibrium and perturbation in five-dimensional guiding-centre space ($\mathbf{x}_G$, $v_{G\|}$, $\mu$) and time. In Eq. (3), $\langle \cdot \rangle = \iint J_v J_x \cdot dv_\| d\mu d\theta d\varphi$, with $J_v$ and $J_x$ being the velocity and space Jacobians. The charge density of guiding-centres $\rho$ is computed as:

$$\rho(\mathbf{x},t) = \frac{1}{n_{e_0}}\sum_s Z_s \int d\mu \mathcal{J}_\mu \cdot \left[\int J_v dv_{G\|}(\bar{F}_s(\mathbf{x},\mathbf{v},t) - \bar{F}_{s,\text{eq}}(r,\theta,v_{G\|}))\right] \quad (6)$$

with $\mathcal{J}_\mu$ the gyro-average operator. Notations are those of ref. [26]. The computational domain extends from inner core ($r/a = 0$) to the material boundaries ($r/a = 1.3$). Flux- or gradient-driven dynamics may be considered. For flux-driven evolution, $\gamma^\kappa = 0$ and the distribution function evolves according to volumetric sources $\mathcal{S}$[21] and penalised heat and momentum sinks $\mathcal{M}^{\text{mat}}(r,\theta)$, $\mathcal{M}^{\text{SOL}}(r)$ and $\mathcal{M}^{\text{wall}}(r)$ that can mimic from poloidally-uniform boundary conditions (Case-2) to the more complex limiter and wall geometries (Case-1). The latter case allows description of the closed to open field lines transition in the SOL. Gradient-driven-like dynamics may also be considered whilst imposing $\mathcal{S} = \mathcal{M}^{\text{mat}}(r,\theta) = \mathcal{M}^{\text{wall}}(r) = 0$. In Case-3, the target distribution function $\mathbb{F}_{F-D}$ is chosen as the statistical distribution at equilibrium from flux-driven Case-2 and the relaxation rate $\gamma^\kappa = 5.43 \, 10^{-5} \sim \gamma_{lin}/10$ is an order of magnitude smaller than the local linear turbulence growth rate $\gamma_{lin}$ at $r/a = 0.7$. Imposing $\mathcal{M}^{\text{SOL}}(r)$ as in Case-2 or cancelling this mask does not alter the dynamics which is dominated by the BGK operator [last term of Eq. (3)], specifically described in ref. [53] and built such as to prevent overdamping zonal modes[11].

Penalisation[54] modifies the equations through introduction of a series of masks $\mathcal{M}^{\text{mat}}(r,\theta)$, $\mathcal{M}^{\text{SOL}}(r)$ and $\mathcal{M}^{\text{wall}}(r)$, combinations of hyperbolic tangents, adjustable in location, shape (for $\mathcal{M}^{\text{mat}}(r,\theta)$), and stiffness. They are illustrated in Fig. 6. Electrons have a Boltzmann response modified by penalisation such that the electric potential $\phi$ in the quasineutrality equation is relaxed towards its expected presheath condition $\Lambda T_e/e$ in the SOL. Additionally, $\phi$ may be biased to $\phi^{\text{bias}}$ in the limiter ($\phi^{\text{bias}} = 0$ in the current study) and is freely evolving elsewhere. $T_e^{\text{b.c.}}$ is the cold electron temperature within limiter and wall, chosen as the minimum $T_e$ value within the computational domain, $\Lambda = -\frac{1}{2}\ln\left[2\pi\frac{m_e}{m_i}\left(1 + \frac{T_i}{T_e}\right)\right]$ and coefficient $\lambda$ (set to unity in the present study) may be used to alter the inertia of the zonal potential. In the gyrokinetic equation, infinite penalisation[55] relaxes $\bar{F}_s$ to a target cold Maxwellian distribution function $\mathbb{G}_{\text{cold}} = n_w(2\pi T_w)^{-3/2}\exp[-(v_{G\|}^2 + \mu B)/2T_w]$, characterised by low wall thermal energy $T_w$ and target density $n_w$. The former is constrained by velocity-space resolution; we typically choose it an order of magnitude lower that temperature at mid radius whilst the target density $n_w$ is chosen so as to ensure particle conservation.

**Physical parameters and robustness**. The reference Tore Supra shot 45511 had 2MW of Ion Cyclotron Resonance Heating on top of 1MW of Ohmic heating injected in a deuterium plasma of normalised size $\rho_\star = \rho_i/a = 1/500$ at mid radius and aspect ratio $a/R_0 = 1/3.3$, $a$ and $R_0$ being, respectively, the minor and major radius. The plasma current is $I_p = 0.8$ MA, the magnetic field on axis is $B_0 = 2.8$T and the density and temperature at mid radius respectively read: $n = 4 \, 10^{19}$ m$^{-3}$ and $T = 0.8$ keV. In GYSELA, a 3MW volumetric heat source comparable in shape to that in the experiment is injected in the central 40% of a deuterium torus of same aspect ratio. Initial density, electron, and ion temperature profiles are the same as in the experiment up to the separatrix. In the core $T_e/T_i > 1$ whilst this ratio reverses in the edge and SOL. To slightly offset the numerical cost of the computations, run on Tier-0 Joliot-Curie at GENCI@CEA and MareNostrum at Barcelona Supercomputing Centre, we assume a reduced magnetic field on axis: $B_0 = 1.7$T, which amounts to computing a plasma column of slightly smaller size $\rho_\star = 1/300$ on a 1/4 wedge torus with $(r, \theta, \varphi, v_\parallel, \mu) = (512, 1024, 64, 128, 64)$ grid. The robustness of the reported main results (observed gradient anisotropy and magnitudes in the limited configuration) are robust whilst varying distribution function initialisation, presheath values in the SOL, penalised temperature $T_w$, limiter shape. A shortfall is robustly found without limiter, even when considering variations in $T_e/T_i$ ratio, safety factor $q$, magnetic shear $s$ and density gradient. These results are reported in detail elsewhere[37].

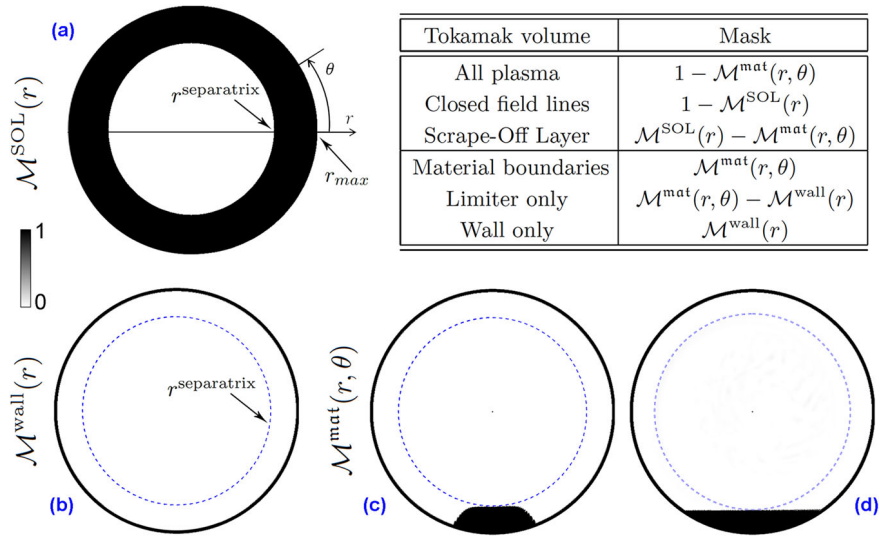

**Fig. 6 The various masks used for penalisation in the gyrokinetic-quasineutrality system.** The resolved domain spans from the very core $r/a = 0$ to the outer wall region at $r/a = r_{max} = 1.3$ (**a**). The outer wall is circular and within $1.25 \leq r/a \leq 1.3$ (**b**). The minimum radius where mask $\mathcal{M}^{mat}(r,\theta) = \frac{1}{2}$ defines the location of the magnetic separatrix at $r/a = r^{separatrix} = 1$. An appealing aspect of penalisation is the ease with which the shape of masks (i.e., the geometry of the material boundaries) can be altered. Two examples are shown in **c**, **d**; all limiter computations shown here are performed in geometry (**d**).

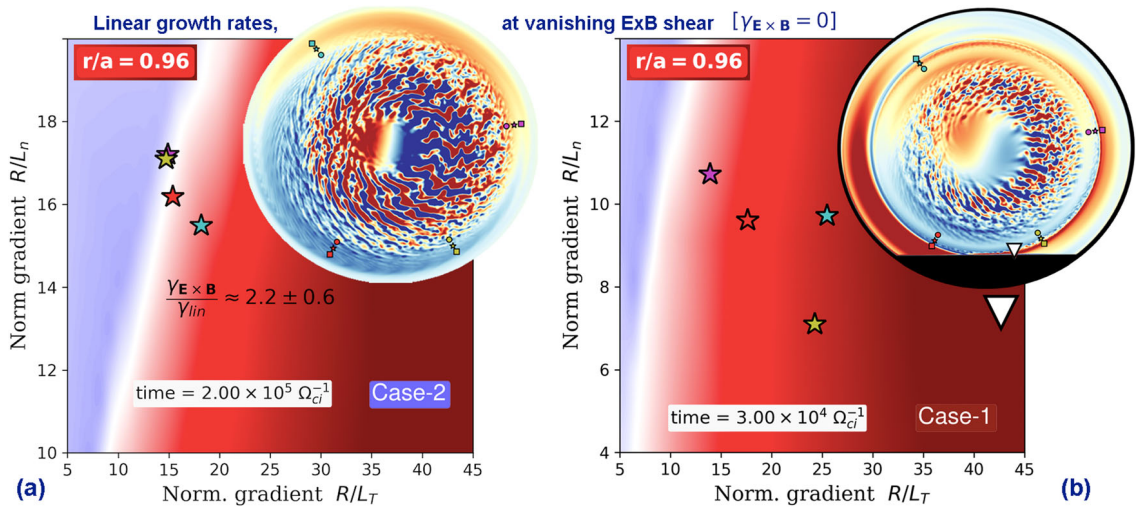

**Fig. 7 Without plasma-wall interplay, the outer edge is linearly stable.** Linear stability analysis is computed at vanishing $\mathbf{E} \times \mathbf{B}$ shear and at radial locations $r/a = 0.9$, 0.96, and 1.02 for the poloidally symmetric flux-driven Case-2 (**a**) and the flux-driven limited Case-1 (**b**). Linear stability of gradient-driven Case-3 is comparable to that of Case-2. Stability at $r/a = 0.96$ is displayed yet all three radial locations provide a qualitatively comparable picture. The following four poloidal locations are shown: $\theta_k = \{9°(\text{magenta}), 126°(\text{cyan}), -118°(\text{red}), -61°(\text{yellow})\}$. A fifth one is shown (**b**) at angle $\theta = -75°$ (white triangle). These locations are also marked in Fig. 1. The outer edge $r/a \geq 0.9$ is linearly stable without limiter (**a**) at all times (shown here at statistical equilibrium ($t\Omega_{ci} = 200,000$)), despite modest core → edge turbulence spreading, illustrated in Fig. 4. Conversely, the outer edge is driven interchange-unstable (**b**) by the presence of the limiter on fast time scales ($t\Omega_{ci} = 30,000$). This drive endures at later times ($t\Omega_{ci} = 250,000$), emphasising robust and steady free energy injection from plasma–boundary interaction.

**Linear stability analysis.** Linear stability analysis of the poloidally symmetric and limited GYSELA profiles is performed using the initial value framework of the GKW code[12], based on the gradient-driven and local (flux-tube) approximations. A typical example is shown in Fig. 7. Growth rates for the most unstable poloidal wavenumbers $k_\theta \rho_i = 0.6$ in poloidally symmetric (Case-2 and Case-3) and limited

(Case-1) configurations are estimated in GKW. The salient conclusions are as follow, with details further provided in ref. [37]: in poloidally symmetric Case-2 and Case-3, GKW finds the edge to be marginally stable at vanishing $\mathbf{E} \times \mathbf{B}$ shear: $\gamma_{lin} \approx 0$. Inclusion of $\mathbf{E} \times \mathbf{B}$ shear $\gamma_E$ would predict the edge to be nonlinearly unconditionally stable past $r/a \geq 0.9$, with $\gamma^{eff} = \gamma_{lin} - \gamma_E < 0$ at all poloidal locations. Analysis with a kinetic trapped electron response provides the same stability assessment for $r/a \geq 0.9$. The situation in Case-1, with the presence of a limiter dramatically changes. A poloidally localised instability at $t\Omega_{ci} = 30,000$ arises about $\theta_k = -61°$(yellow), dominantly of interchange character whilst the midplane $\theta_k = 9°$ remains linearly stable. Further details are given in ref. [37].

**Causal inference.** The following vorticity equation can be inferred from the primitive gyrokinetic equations including $\mathbf{E} \times \mathbf{B}$ drift and FLR at leading order:

$$\partial_t \langle \Omega_r \rangle + \frac{1}{r} \partial_r r \left[ \langle v_{Er} \Omega_r \rangle + \langle v_{\star r} \Omega_r \rangle - \left\langle v_{\star\theta} \frac{1}{r} \partial_\theta E_r \right\rangle \right] = r.h.s \quad (7)$$

$$r.h.s \approx -\partial_t \langle \Omega_\theta \rangle - \frac{1}{r} \partial_\theta \langle (v_{E\theta} + v_{*\theta}) \Omega_\theta \rangle - \partial_r \frac{1}{2r} \partial_\theta \langle v_{E\theta}^2 \rangle + \frac{1}{2r^3} \partial_\theta \partial_r \langle r^2 v_{Er}^2 \rangle$$
$$- \frac{1}{r} \partial_\theta \left\langle v_{*r} \frac{1}{r} \partial_\theta v_{E\theta} \right\rangle \quad (8)$$

$$\Omega_r = \partial_r(r\partial_r \phi)/r \quad \& \quad \Omega_\theta = \partial_\theta^2 \phi / r^2 \quad (9)$$

$$v_{Er} = -\partial_\theta \phi / r \quad \& \quad v_{E\theta} = \partial_r \phi = -E_r \quad (10)$$

$$v_{*r} = -\partial_\theta p_\perp / r \quad \& \quad v_{*\theta} = \partial_r p_\perp \quad (11)$$

where $\langle \cdot \rangle$ denotes an average over toroidal angle $\varphi$. The net Transfer Entropy $\Delta_{X,Y}(TE)[k] = TE_{Y\to X}[k] - TE_{X\to Y}[k]$ is a measure of net flow of information between processes $X$ and $Y$, at timelag $k$. This measure has been systematically applied to the possible permutations of quantities in Eq. (7) and in this manuscript to the following set:

$$(X, Y) \in \left\{ \langle \Omega_r \rangle, \langle v_{Er} \Omega_r \rangle, \langle v_{*r} \Omega_r \rangle, -\left\langle v_{*\theta} \frac{1}{r} \partial_\theta E_r \right\rangle \right\} \quad (12)$$

where the TE from process $Y$ to process $X$ is defined as:

$$TE_{Y\to X}(k) = \sum p(x_{n+1}, x_{n-k}, y_{n-k}) \log \left( \frac{p(x_{n+1}|x_{n-k}, y_{n-k})}{p(x_{n+1}|x_{n-k})} \right) \quad (13)$$

and $(x_i)$ and $(y_i)$ respectively denote time series of realisations of $X$ and $Y$, with $0 \le i \le n$.

### Data availability
Raw data generated by the GYSELA code is available from the corresponding author upon reasonable request.

### Code availability
The GYSELA suite of codes used for this article is available upon request via the GitHub repository https://gyselax.github.io.

## Acknowledgements
The authors acknowledge stimulating discussions with P.H. Diamond and participants at the 2019 and 2021 "Festival de Théorie" in Aix-en-Provence. This work has been carried out within the framework of the EUROfusion Consortium and was supported by the EUROfusion Theory and Advanced Simulation Coordination (E-TASC) initiative under the TSVV (Theory, Simulation, Verification, and Validation) "L-H transition and pedestal physics" project (2019-2020) and TSVV "Plasma Particle/Heat Exhaust: Gyrokinetic/Kinetic Edge Codes" (2021–2025). It has received funding from the Euratom research and training programme 2014–2018 and 2019–2020 under grant agreement No 633053. The authors gratefully acknowledge funding from the European Commission Horizon 2020 research and innovation programme under Grant Agreement no. 824158 (EoCoE-II). The views and opinions expressed herein do not necessarily reflect those of the European Commission. This research was supported in part by the National Science Foundation under Grant no. NSF PHY-1748958. We acknowledge PRACE for awarding us access to Joliot-Curie at GENCI@CEA, France, and MareNostrum at Barcelona Supercomputing Centre (BSC), Spain.

## Author contributions
G.D.-P., Ph.Gh., and E.C. designed and implemented the penalisation technique, G.D.-P. performed the GYSELA computations, wrote the Transfer Entropy method, and performed the analysis. V.G. is the main developer of the the GYSELA code. P.D. implemented the collision operator. G.D-P., Ph.Gh., Y.M., and Y.S. discussed the sheath-related issues. L.V and F.C. provided the experimental data. F.W. and Y.C. performed the linear GKW analysis. G.D-P., Y.S., X.G., and Ph.Gh. derived and discussed the vorticity equation. G.D-P. wrote the manuscript. All authors discussed the results and commented on the paper.

## Competing interests
The authors declare no competing interests.

## Additional information
**Supplementary information** The online version contains supplementary material available at https://doi.org/10.1038/s42005-022-01004-z.

**Correspondence** and requests for materials should be addressed to Guilhem Dif-Pradalier.

**Peer review information** *Communications Physics* thanks István Cziegler and the other, anonymous, reviewer(s) for their contribution to the peer review of this work. This article has been peer reviewed as part of Springer Nature's Guided Open Access initiative.

**Reprints and permission information** is available at http://www.nature.com/reprints

**Publisher's note** Springer Nature remains neutral with regard to jurisdictional claims in published maps and institutional affiliations.